\documentclass[twocolumn,twoside,slac_two,floatfix]{revtex4}
\usepackage{graphicx,amsmath}
\usepackage{fancyhdr}
\begin{document}

\title{On the correct intrinsic VHE properties of the BL Lac H\,2356-309}

%

\author{L. Costamante}
\affiliation{HEPL/KIPAC, Stanford University, Stanford, CA 94024, USA \\
Now at: Dept. of Physics, Universit\'a degli Studi di Perugia, I-06123 Perugia, Italy}

\begin{abstract}
The high-energy-peaked BL Lac H\,2356-309 (z=0.165) was detected by HESS
at very high energies (VHE, $\gtrsim100$ GeV) with relatively high significance
in the years 2004-2007, allowing a good determination of its gamma-ray spectrum.
After correction for the interaction with the diffuse extragalactic background light (EBL),
the VHE spectrum is  flat ($\Gamma\sim1.9-2$) over a decade in energy, locating 
the gamma-ray peak around or above 0.6-1 TeV.
This is remarkably at odds with the interpretation and modeling provided by HESS, 
which do not correspond to the source properties and  can be excluded with high confidence.
The overall GeV-to-TeV characteristics of H\,2356-309 seem intermediate between
the ``TeV-peaked" (Fermi-faint) and ``100 GeV-peaked" (Fermi-bright) BL Lac objects,
and difficult to reconcile with the shape of the synchrotron emission
in a single-zone SSC scenario.
\end{abstract}

\maketitle

\thispagestyle{fancy}

\section{Introduction}
The BL Lac object H\,2356-309 (z=0.165) is a High-frequency-peaked 
BL Lac (HBL), also called High-Synchrotron-Peaked blazar (HSP, \citep{2lac}).
It is usually bright in the X-ray band, and during BeppoSAX observations in 1998
it presented a synchrotron peak in the spectral energy distribution (SED) 
above few keV, which is the defining characteristic
of the so-called ``extreme BL Lacs" \citep{extreme}.

At VHE, H\,2356-309  was detected for the first time in 2004  by HESS 
\citep{nature,hess2356}. The VHE spectrum was found to be significantly harder than expected
from a source at that redshift, considering the softening 
effects of $\gamma$-$\gamma$ absorption on  the 
diffuse extragalactic  background light (EBL), in the detected energy range. 
Together with other even harder sources, this fact lead to the discovery 
of a low intensity of the EBL in the component produced by the direct starlight \citep{nature}.

Since 2004, H\,2356-309 has been monitored by HESS for several years, 
reaching a total detection of $\sim$13 $\sigma$ in the timespan 2004-2007,
with  an average flux at the level of $\sim$1.6\% of the Crab flux (above 240 GeV).
In this epoch, three simultaneous multi-wavelength campaigns were performed, in
2004 with RXTE and 2005 with XMM-Newton \citep{hess2356,icrc08}, which allowed
the characterization of its SED.
The results of these campaigns, the VHE monitoring and
a synchrotron self-Compton (SSC) modeling of the data were published
by the HESS Collaboration in Abramowski et al. 2010 \citep{2356mwl}.

In that paper, however, while all the data analysis is correct, 
the HESS Collaboration has apparently misinterpreted the VHE data, 
providing a wrong assessment of the intrinsic VHE properties of the source.
It is also not consistent with previous publications on the same results.
This contribution presents the arguments against the interpretation in \citep{2356mwl}
and tries to provide a more accurate description of the actual gamma-ray properties 
(from the GeV to the TeV band) of this BL Lac object.

\section{Wrong assessment of the intrinsic VHE properties by HESS}
\begin{figure}
\includegraphics[angle=-90,width=85mm]{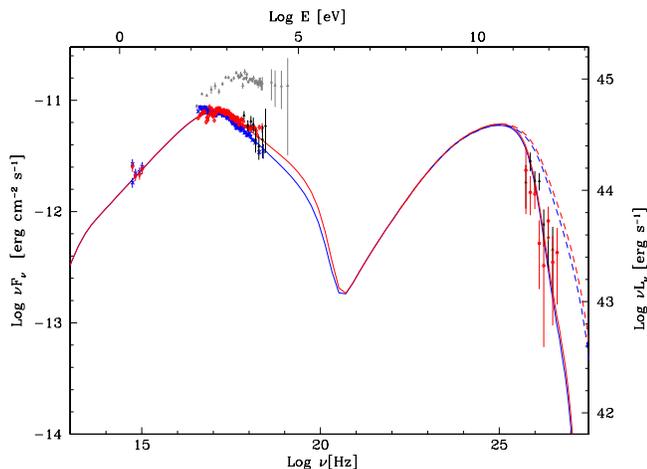}
\caption{The SED of H\,2356-309 in different epochs, from the HESS paper (Fig. 8 in \citep{2356mwl}).
The XMM observations in 2005 are plotted in color (blue for June 13, red for June 15), 
while the RXTE 2004 data are plotted as black triangles.
In the VHE band, the observed (i.e. EBL-absorbed) HESS data are plotted, as black triangles
and red dots for the years 2004 and 2005, respectively.
The curves are the HESS modeling of the 2005 SEDs with a single-zone SSC model, 
as described in \citep{2356mwl}, with (solid lines) and without (dashed lines) the EBL effects 
included \citep{franceschini}.
The SSC model for the 2004 data (RXTE and HESS) is identical to the fit 
to the June 15, 2005 data (red curve).
}
\label{f1}
\end{figure}

Fig. \ref{f1} shows the overall SED of H\,2356-309
during the HESS multiwavelength campaigns in 2004 and 2005 (from \citep{2356mwl}). 
The two lines shows the single-zone SSC modeling proposed by the HESS Collaboration,
with the dashed lines representing the source intrinsic emission (i.e. before absorption effects on the
EBL calculated according to \cite{franceschini}).
This SSC model is claimed by HESS to represent well the gamma-ray SED properties, 
with a Compton peak  around 50-100 GeV and a steep intrinsic VHE spectrum:
the local slope of the intrinsic SSC model (dashed line) in the detected VHE 
range is $\Gamma_{int}\approx2.65-2.7$.

\begin{figure}
\includegraphics[width=85mm]{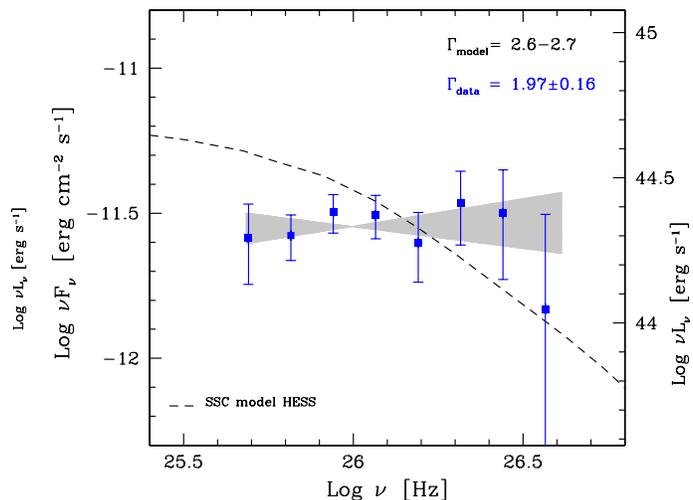}
\caption{Zoom of the SED in the VHE band: the average 2004-2006 HESS data are plotted, 
this time corrected for absorption with the EBL \citep{franceschini}.
The dashed line correspond to the SSC model by HESS 
(the red dashed curve in Fig. \ref{f1}). The discrepancy is clearly visible:
the model photon index is around 2.65-2.7 in the HESS energy band, while the data photon
index is $\Gamma=1.97\pm0.16$ (statistical error only, shown by the grey butterfly).
The model slope is excluded by the data at 99.99\% confidence level.
}
\label{f2}
\end{figure}

However, Fig \ref{f2} shows this not to be the case.   
Accounting for EBL absorption  --either by correcting the observed data points, as done here, or by fitting 
an absorbed model-- the intrinsic spectrum is in fact much harder. 
Fig. \ref{f2} presents the overall 2004-2006 average HESS data from \citep{2356mwl}, corrected for EBL absorption 
with the same EBL calculation \citep{franceschini}.  
Using a power-law model,  the intrinsic photon index of the absorption-corrected data 
is $\Gamma_{int}=1.97\pm0.16$ (statistical error only), or $\Gamma_{int}=1.91\pm0.18$ excluding 
the last point (which is more like an upper limit, see \citep{2356mwl}).
The discrepancy with the slope of the SSC model is large and highly significant: 
the $\Delta\chi^2$ needed to recover an index  of 2.65 is 15, corresponding to a probability 
of $\sim$1E-4 (for 1 parameter).    The HESS SSC model, therefore, is excluded by the data at 
$\sim$99.99\% confidence level.
 
Note that the 2004-2006 overall spectrum is mostly dominated --90\% of the excess signal-- by 
the 2004-2005 data set. Moreover, the spectrum in 2006 seems to have been steeper than in 2004-2005 
(see Table 4 in \cite{2356mwl}), thus the 2004-2005 spectrum should actually be slightly harder than 
the shown average, increasing the discrepancy.

The importance of this discrepancy is not merely quantitative, but qualitatively:
it changes dramatically the luminosity and the location of the gamma-ray peak. 
The latter is in fact at energies one-to-two orders of magnitude higher than
claimed by HESS, and at lower apparent fluxes (see Fig. \ref{f3}). 
This changes completely the physical parameters and the character of the SED.

It cannot therefore be accepted the explanation in the HESS paper about the ``slight difference 
between model and data" as  due to ``the inclusion of all the multiwavelength 
data in the curved SSC fits", because therein lies the core of the problem: 
precisely because constrained by the observed shape of the synchrotron emission, 
a one-zone SSC model (with these parameters) is NOT able to reproduce 
the intrinsically hard HESS data.


\subsection{Further arguments}
The VHE data shown in Fig. \ref{f1} are not the overall average, but correspond to the two years 
2004 and 2005 considered separately.  It could be argued that the HESS SSC modeling aims to reproduce 
these two single epochs, and that the lower statistics of each single dataset makes the model viable, 
despite being excluded by the sum.
However, this is not a valid argument: since the model gives identical slopes/characteristics
in the two years,  
if it represented well the spectrum in both epochs  %
it should fit well also the sum of the two, by definition. %
The situation is similar to having a constant source, properties-wise:
the accumulation of statistics is revealing if the model is compatible or not with the data,
and in this case it is not.

If this SSC modeling reproduced the actual source spectrum in just one of the two epochs,
the spectrum in the other epoch should be much harder than the average, to compensate, 
and this should show in the data (as well as in the models, i.e. the SSC models for 2004 and 2005 
should be very different from each other, to be consistent with the average).
Again this seems not the case: the measured slopes of the 2004 and 2005 data sets are identical 
within the errors ($\Gamma_{\rm VHE}=2.97\pm0.19$ vs  $2.99\pm0.39$ respectively, 
see Table 4 in \cite{2356mwl}).

In other words, there is no way that the HESS modeling can represent well the HESS data.
Simply stated, the proposed SSC model does not correspond to the SED of H\,2356-309,
and thus the physical analysis based on its parameters should be discarded.

\section{The problem of plotting VHE data}
There is a general issue in the interpretation of the VHE data.  
When presenting the SED of TeV blazars, it is becoming customary to plot
only the observed (i.e. not EBL-corrected) data points and  to absorb
the possible model/curve, plotting at most the curves before EBL absorption.
{\it This is often misleading and a dangerous habit, and should be avoided.}

Even if in principle it is the most proper method for a statistical analysis of the data 
and to derive fitting parameters, for plotting purposes it does not accurately
display the source SED features. It misleads the viewers  on the true properties 
of the object by mixing them with those of intervening systems.
It also creates confusion when trying to find an adequate set of parameters 
with an emission model, because all curves seem  to fit well the SED data  
when spectra are very steep.

The alternative approach is to correct the data points for the EBL 
(e.g. with the optical depth calculated at the average photon energy in the bin, 
which provides the same fit parameters of absorbing the model; 
see e.g. \citep{nature}), and plotting the EBL de-absorbed data in the SED.
Both methods, if done correctly, should and do give the same results.
This second approach, however, provides visually the most accurate representation of the true 
properties of the source  (which is the focus of a SED study), 
without confusion with those of intervening systems. 

Furthermore, this method achieves consistency with the conventions used 
in the other energy bands on the SED: 
in the Optical-UV bands, data are usually plotted after correction for at least
Galactic extinction.  In the X-ray band, data/slopes are always shown with correction for
Galactic or line-of-sight intervening column densities. 
The reason is precisely to focus on the source physics, if that is the goal of the study.
Otherwise any extragalactic object would always appear with fake peaks in the SED 
typically at ~1-2 keV (or more, depending on the column density) 
and in the red-infrared band.

In conclusion, EBL-absorption should be treated and accounted for  in the SED
as any other absorption effect from intervening systems in the electromagnetic spectrum.

\begin{figure}
\includegraphics[width=85mm]{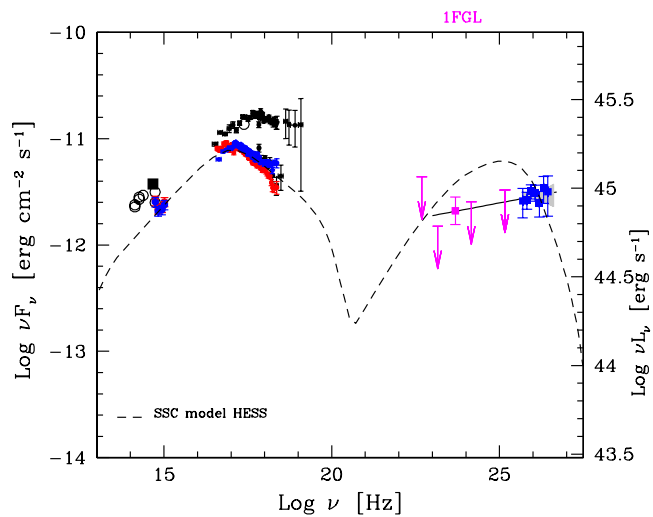}
\caption{The updated (correct) SED of H\,2356-309. Black markers represent historical data
(including 1998 BeppoSAX and 2004 RXTE data). The XMM-Newton 2005 data (blue and red circles), 
re-analyzed with SAS v11 using the same procedures described in \citep{2356mwl}, 
are shown as in Fig. \ref{f1}.
In gamma-rays, the HESS data (blue squares) are shown as in Fig. \ref{f2}. 
The LAT data (magenta point and upper limits) are from the 1-FGL catalog \citep{1fgl}. 
Data obtained using the web tools of the ASI Science Data Center.}
\label{f3}
\end{figure}

\section{\label{sed}The correct SED of H\,2356-309}
The real intrinsic properties of H\,2356-309, as given by the HESS data, 
are shown in Fig. \ref{f3}. These are: 
\begin{enumerate}
\item a flat VHE spectrum of photon index $\Gamma_{int}\approx$1.9-2 over the whole HESS range;
\item a Compton peak either around $\sim$600 GeV (as determined by a log-parabolic fit 
of the HESS data), or $\gtrsim$1 TeV (assuming the single power-law model). 
The statistics of the data do not allow yet  to distinguish between the two cases
(the curvature parameter can be zero within the errors).
\end{enumerate}

Quite interestingly, the extrapolation of the HESS 2004-2006 average spectrum 
to the Fermi-LAT band corresponds  almost "spot on" to the fluxes in the 1FGL catalog 
(2008-2009, TS=50, detection mainly in the 1-3 GeV bin; \citep{1fgl}).
A single power-law model over 3 decades in energy provides surprisingly a good fit ($\chi^2_r\sim0.3$)
with $\Gamma_{int}=1.94\pm0.03$.
The 2FGL values, instead, show a reduced overall flux (by roughly 2x), but retaining 
very similar spectral properties ($\Gamma=1.89\pm0.17$, \citep{2fgl}).

With the big caveats of the non-simultaneity of the gamma-ray data and of the different integration 
times of the Optical-to-X-ray data vs the gamma-ray data, such a Gev-TeV gamma-ray spectrum is not easy 
to model  within one-zone SSC scenarios, given the very different  synchrotron spectra 
in the optical to X-ray bands, which trace directly the shape of the electron distribution.
Regardless, the average SED seems synchrotron-dominated in luminosity, and thus 
the electron cooling as well.

\begin{figure}
\includegraphics[width=80mm]{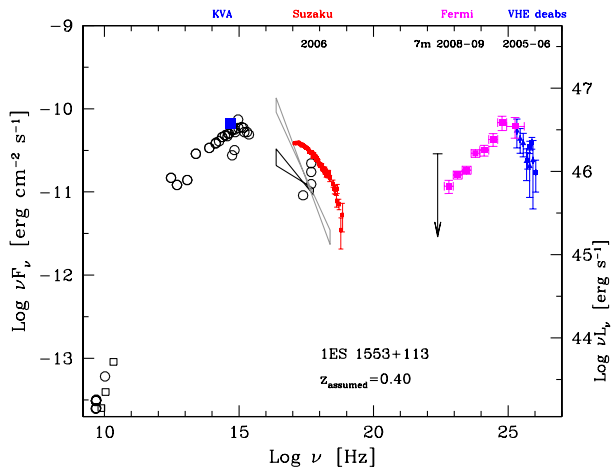}
\caption{Typical SED of a Fermi-bright ``100-GeV peaked" HBL:  1ES\,1553+113.
They are characterized by soft intrinsic VHE spectra locating the gamma-ray peak 
close to 100 GeV  (like PKS 2005-489 or PKS 2155-304).  
The redshift is constrained by HST/COS data \citep{danforth}.
Fermi-LAT spectrum from \citep{lat1553}.  Other data from \citep{reimer1553}.
VHE data corrected for EBL according to \citep{franceschini}.
}
\label{f4}
\end{figure}
\begin{figure}
\includegraphics[width=80mm]{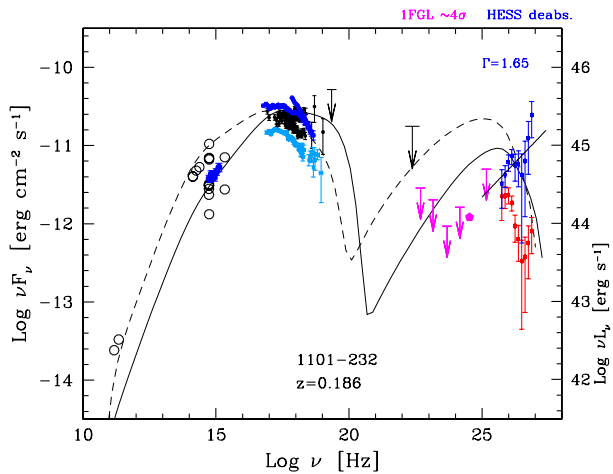}
\caption{Typical SED of a Fermi-faint TeV-peaked HBL: 1ES\,1101-232.
This was the first HBL object discovered with a gamma-ray peak above few TeV
(i.e. with a hard VHE spectrum irrespective of the EBL model used; \citep{nature}).
Fermi-LAT data from the 1-FGL catalog \citep{1fgl}. The LAT point corresponds to a flux estimate 
for the $\sim4$ $\sigma$ signal in the 1-FGL catalog. Other multiwavelength data/campaigns from
\citep{reimer1553} and refs therein. VHE observed data (red) corrected for absorption (blue) 
according to \citep{franceschini}.}
\label{f5}
\end{figure}

\section{Comparison with other HBL}
HBL objects are typically characterized by hard ($\Gamma<2$) Fermi-LAT spectra \citep{2lac}
but show a wide range of VHE slopes, depending on the location of the gamma-ray SED peak.
To this respect,  H\,2356-309 seems intermediate between two types of HBL objects:  
those with the peak around $\sim100$ GeV, i.e. inbetween the Fermi-LAT and VHE bands
(see for example Fig. \ref{f4}),
and those with a gamma-ray peak above 1 TeV, or ``TeV-peaked" HBL (Fig. \ref{f5}).
The formers are characterized by soft intrinsic VHE spectra,  
and are more easily detected in Fermi because the peak of the SED 
is close to the Fermi-LAT band. 
The TeV-peaked HBL, instead, are characterized by hard intrinsic VHE spectra 
($\Gamma_{\rm VHE}<2$), 
irrespective of the EBL model, locating the peak all the way beyond the detected VHE range.
They remain often very weak or undetected in Fermi,
because the LAT band falls now deep into the valley between the sycnhrotron and gamma-ray humps.
These objects however, despite being rarer among HBLs, are at present the most challenging AGN 
for SSC modeling and to gain new insigths on particle acceleration and emission mechanisms.
(e.g. \citep{katar,lefa,tav0229,olga}).
%

\bigskip 

\noindent {\bf  Disclaimer:}
{\small As co-author of the HESS paper on H\,2356-309, 
these findings were already provided to the HESS Collaboration, in the early stage and also after 
publication, to no positive effect. The co-authorship, therefore,
is to be intended as data-related  only, as proposer and PI of the multi-wavelength campaigns 
and for providing the XMM-Newton data analysis. 
It does not extend to the interpretation, which was finalized after 
I left the collaboration.}

\bigskip 


\end{document}